
\def\IR{{\hbox{{\rm I}\kern-.2em\hbox{\rm R}}}}
\def\IB{{\hbox{{\rm I}\kern-.2em\hbox{\rm B}}}}
\def\IN{{\hbox{{\rm I}\kern-.2em\hbox{\rm N}}}}
\def\IC{{\ \hbox{{\rm I}\kern-.6em\hbox{\bf C}}}}

\def\IZ{{\hbox{{\rm Z}\kern-.4em\hbox{\rm Z}}}}
\def\to{\rightarrow}
\def\underarrow#1{\vbox{\ialign{##\crcr$\hfil\displaystyle
{#1}\hfil$\crcr\noalign{\kern1pt
\nointerlineskip}$\longrightarrow$\crcr}}}
%
\def\d{{\rm d}}

\def\tilde{\widetilde}
\input phyzzx
\overfullrule=0pt
\tolerance=5000
\overfullrule=0pt
\twelvepoint

\twelvepoint
\pubnum{IASSNS-HEP-92/38}
\date{June, 1992}
\titlepage
\title{A CLOSED, EXPANDING UNIVERSE IN STRING THEORY}
\vglue-.25in
\author{Chiara R. Nappi\foot{Research supported in part by
the Ambrose Monell Foundation.} and Edward Witten
\foot{Research supported in part by NSF Grant
PHY91-06210.}}
\medskip
\address{School of Natural Sciences
\break Institute for Advanced Study
\break Olden Lane
\break Princeton, NJ 08540}
\bigskip
\abstract{We present a conformal field theory -- obtained from a gauged
WZW model -- that describes   a closed, inhomogeneous expanding
and recollapsing universe in $3+1$ dimensions.
A possible violation of cosmic censorship
is avoided because the universe recollapses just when a naked singularity
was about to form.
The model has been chosen to have
$c=4$ (or $\widehat c=4$ in the supersymmetric
case), just like four dimensional Minkowski space.}
\endpage

\REF\bars{I. Bars, Nucl. Phys. {\bf B334} (1990) 125; I. Bars and
D. Nemeschansky, Nucl. Phys. {\bf B348} (1991) 89.}
\REF\bardakci{K. Bardakci, E. Rabinovici, and B. Saring, Nucl. Phys.
{\bf B299} (1988) 151; A. Altschuler, K. Bardakci, and E. Rabinovici,
Comm. Math. Phys. {\bf B118} (1988) 241. }
\REF\witten{E. Witten, ``String Theory And Black Holes,'' Phys. Rev.
{\bf D44} (1991) 314. }
\REF\gins{P. Ginsparg and F. Quevedo, ``Strings On Curved Space-Times:
Black Holes, Torsion, and Duality'' LA-UR-92-640.}
\REF\ish{N. Ishibashi, M. Li, and A. R. Steif, Phys. Rev. Lett. {\bf 67}
(1991) 3336.}
\REF\horne{J. H. Horne, G. T. Horowitz, and A. R. Stief, Santa Barbara
preprint UCSB-TH-91-53.}
\REF\horo{J. H. Horne and G. T. Horowitz, Nucl. Phys. B {\bf 368} (1192) 444.}
\REF\kar{S. K. Kar, S. P. Khastgar, and G. Sengupta, IP/BBSR92-35.}
\REF\mah{S. Mahapatra, Tata Institute preprint (May, 1992).}
\REF\stets{I. Bars and K. Sfetsos, ``Global Analysis Of New Gravitational
Singularities In String And Particle Theories'' USC-92/HEP-B1.}
\REF\hor{P. Horava, ``Some Exact Solutions Of String Theory In Four
And Five Dimensions,'' Phys. Lett. B {\bf278} (1992) 101.}
\REF\gersh{D. Gershon, ``Exact Solutions Of Four-Dimensional Black Holes
In String THeory'' TAUP-1937-91.}
\REF\koun{C. Kounnas and D. Lust, ``Cosmological String Backgrounds
{}From Gauged WZW-models'' CERN-TH-6494-92.}
Gauged WZW models are one of the few techniques we know for constructing
conformal field theories that describe
time dependent solutions of string theories.
This approach was considered by Bars [\bars]; the $SU(2)/U(1)$
model was studied by Bardakci et. al. to get a sigma model
approach to parafermions [\bardakci]; one of us studied the $SL(2,\IR)/U(1)$
model and showed that it described a two dimensional black hole
[\witten].  Since then, there have been many other papers [\gins--\koun].
\foot{In particular, [\hor] and [\gersh] discuss different real
forms of the coset we will analyze, not directly leading to the cosmological
solution.  This coset has also been mentioned by several other
authors without analyzing its content.}

The two dimensional black hole is a solution of $D=2$ string
theory.   Clearly, it is of interest to find a conformal field theories
that describe alternative solutions of the same string theories that
are used to build more or less realistic models of particle physics.
In applying string theory to the real world, one represents the
vacuum by $M^4\times K$, where $M^4$ is four dimensional Minkowski space
represented by a free field theory with $c=4$ (or $\widehat c=4$,
supersymmetrically) and $K$ is some internal space represented by
a conformal field theory of appropriate central charge.  To find
a realistic cosmological solution, it suffices to find conformal field
theory with four dimensional target space $N$ and $c=4$, so that
$M^4\times K$ can be replaced by $N\times K$.

Moreover, though the possible formation of
singularities is one of the questions of interest, we do not want
the singularities to be ``built in'' in the initial conditions.
We want $N$ to admit a smooth, complete
initial value hypersurface.

In this paper, we will describe one
example satisfying these conditions.
The model turns out to describe an anisotropic expanding and
contracting closed universe.
The topology of a spatial slice is ${\bf S}^3$, the
three-sphere.  The three-sphere expands from an initial singularity
at which the metric vanishes in one of the three directions and the
volume therefore vanishes, and collapses to a final singularity in which
the metric vanishes in a different
one of the three directions, so that the volume
again vanishes. The recollapse occurs just at the moment that a naked
 singularity was about to form.

We take the target space to be $G=SL(2,\IR)\times SU(2)$, of dimension six.
For a WZW model at levels $(k',k)$, the central charge would be
$$c_0= {3k'\over k'+2}+{3k\over k+2}. \eqn\jjd$$
We will take the gauge group  $H$ to be a two
dimensional abelian subgroup of $G_L\times G_R$,
so as to end up with a four dimensional model.
Gauging $H$ will reduce the central charge by two, so we want
$c_0=6$.  This gives the relation between $k$ and $k'$ that we wish to
impose.  However, in practice all we will do with the model in this
paper is to determine the space-time geometry in the field theory
limit of astronomically large $k$ and $k'$.  To this end,
the condition $c_0=6$ can be reduced to $k'=-k$, which will give
some minor simplifications in the formulas.

$H$ must be chosen to respect a condition of anomaly cancellation,
or the gauged WZW model will not exist.
This condition is as follows.  Let $\Tr$ be the trace on the $SL(2,\IR)
\times SU(2)$ Lie algebra given by $\Tr=k'\tr_1+k\tr_2$, where $\tr_1$
and $\tr_2$ are the traces in the two dimensional representations of
$SL(2,\IR)$ and $SU(2)$.  Let $T_a, \,\,a=1,2$ be the generators of
 $H$, represented in $G_L\times G_R$ by matrices $(T_{a,L},T_{a,R}),
\,\,a=1,2$.  Then  the standard condition of anomaly cancellation is
$$ \Tr T_{a,L}T_{b,L}=\Tr T_{a,R}T_{b,R},~~~~{\rm for}~a,b=1,2. \eqn\momn$$

We also wish to pick $H$ so
as to end up with a target space of signature $-+++$.
A suitable $H$, isomorphic to $\IR\times \IR$, is generated infinitesimally
by the following transformation of $(g_1,g_2)\in SL(2,\IR)\times SU(2)$:
$$\eqalign{
\delta g_1 = &\epsilon \sigma_3 g_1 + ({\widetilde \epsilon}\cos\alpha
+\epsilon \sin\alpha) g_1 \sigma_3 \cr
\delta g_2 = &i{\widetilde
\epsilon}\sigma_2 g_2 + i(-{\widetilde\epsilon}\sin\alpha +
 \epsilon \cos\alpha) g_2 \sigma_2.\cr} \eqn\hhh$$
Here $\epsilon$ and $\tilde \epsilon$ are infinitesimal parameters, $\alpha$
is a free parameter that enters in defining the model,
and $\sigma_i,\,\,i=1\dots 3$, are the standard $2\times 2$ Pauli
matrices,
$$ \sigma_1 = \pmatrix{0&1\cr 1&0\cr}\qquad  \sigma_2
= \pmatrix{0&-i\cr i&0\cr} \qquad  \sigma_3
= \pmatrix{1&0 \cr 0 &-1\cr}.\eqn\rogu$$
\hhh\ characterizes an anomaly-free subgroup in the case $k=-k'$.
If we set $c_0=6$, we would have to modify \rogu\ by terms that
vanish for $k,k'\to\infty$ and so do not affect the classical
limit of the geometry.

\def\d{{\rm d}}
The standard WZW action is
$$S_{WZW}(g) = -{1\over 4\pi}\int_\Sigma \d^2z
\tr[g^{-1}\partial g g^{-1}
{\bar\partial}g] -  \Gamma (g)\eqn\uno$$
where
$$\Gamma (g) = {i\over 12\pi}\int_B
\tr[g^{-1}\d g\wedge g^{-1} \d g\wedge g^{-1}\d g]. \eqn\wzw$$
Above $\Sigma$ is a Riemann surface, $B$ a three-manifold with
$\partial B=\Sigma$, and $g$ may be valued in either $SL(2,\IR)$
or $SU(2)$.
To gauge the symmetry introduced above, we introduce abelian
gauge fields $A$, $\tilde A$, with
$$\delta A_i = -\partial_i\epsilon \qquad
\delta{\widetilde A_i}= - \partial_i {\widetilde \epsilon}.
\eqn\mcmc$$
For $k'=-k$, the gauge invariant Lagrangian turns out to be
$$\eqalign{L(g_1,g_2,A,\widetilde A) &=-k S_{WZW}(g_1)+kS_{WZW}(g_2)\cr
&+{k\over 2\pi}\int \d^2z\left( A_z \tr[{\bar\partial }g_1g_1^{-1}\sigma_3]
+({\widetilde A_{\bar z}}\cos\alpha + A_{\bar z}\sin\alpha)
\tr[g_1^{-1} \partial g_1 \sigma_3]\right)\cr
&-{ik\over 2\pi}\int \d^2z\left({\widetilde A_z} \tr[\sigma_2
{\bar\partial} g_2 g_2^{-1}]
+ (A_{\bar z}\cos\alpha - {\widetilde A_{\bar z}}\sin\alpha)
\tr[\sigma_2g_2^{-1}\partial g_2]\right)\cr
&+{k\over 2\pi}\int \d^2z (A_z{\widetilde A_{\bar z}}\cos\alpha +
A_z A_{\bar z}\sin\alpha)\tr[\sigma_3g_1\sigma_3g_1^{-1}]\cr &
+{k\over 2\pi}\int \d^2z ({\widetilde A_z}A_{\bar z}\cos\alpha
- {\widetilde A_z} {\widetilde A_{\bar z}}\sin \alpha) \tr[\sigma_2 g_2
\sigma_2 g_2^{-1}]
\cr &+{k\over \pi}\int \d^2z(A_zA_{\bar z} + {\widetilde A_{\bar z}}
{\widetilde A_z}).  \cr} \eqn\two$$

Now we want to pick a gauge condition, integrate out the gauge fields,
and reduce to a sigma model with a four dimensional target space.
In a suitable open set in field space, the gauge can be uniquely
fixed by requiring $g_1$ to be of the form
$$ g_1 = \pmatrix{\cos \psi & \sin \psi\cr -\sin \psi &\cos \psi \cr}\eqn\idi$$
This gauge condition can be imposed precisely in that region of field
space in which
$$ |\tr \sigma_3 g_1 \sigma_3 g_1{}^{-1}|\leq 2. \eqn\kdid$$
This region of field space will lead to our cosmological
model.  The region with $|\tr \sigma_3 g_1 \sigma_3 g_1{}^{-1}|>2$ can also
be considered (with a different gauge condition), but leads to a model
with closed timelike curves, as we will see later.

In this gauge, the above lagrangian becomes
$$\eqalign{S = & -{k\over 2\pi}\int \d^2z \partial \psi{\bar\partial \psi}
 -{k\over 4\pi}
\int \d^2z  \tr[g_2^{-1}\partial g_2 g_2^{-1}{\bar\partial g_2}]
-k\Gamma (g_2)\cr
&-i{k\over 2\pi}\int \d^2z\{{\widetilde A_z} \tr[
\sigma_2{\bar\partial} g_2 g_2^{-1}]
+ (A_{\bar z}\cos\alpha - {\widetilde A_{\bar z}}\sin\alpha)
 \tr[\sigma_2g_2^{-1}\partial g_2]\}\cr
&+{k\over \pi}\int \d^2z \cos 2\psi (A_z{\tilde A_{\bar z}}\cos\alpha
+ A_zA_{\bar z}\sin\alpha)\cr & +{k\over 2\pi}
\int \d^2z ({\tilde A_z} A_{\bar z}\cos\alpha -
{\tilde A_z}{\tilde A_{\bar z}}\sin\alpha)\tr[\sigma_2g_2\sigma_2g_2^{-1}]
+{k\over \pi}\int \d^2z (A_zA_{\bar z} + {\tilde A_{\bar z}}{\tilde A_z}).\cr}
 \eqn\three$$
This action is quadratic in the gauge fields, which therefore can be
integrated out. The answer is
$$\eqalign{&S =  -{k\over 2\pi}\int \d^2z \partial \psi{\bar\partial \psi}
 -{k\over 4\pi}
\int \d^2z \tr[g_2^{-1}\partial g_2 g_2^{-1}{\bar\partial g_2}]
-k\Gamma (g_2)\cr
&-{k\over 4\pi}\int \d^2z  {{\tr[\sigma_2{\bar\partial} g_2 g_2^{-1}]
({\cos 2\psi +\sin\alpha})\tr[\sigma_2g_2^{-1}\partial g_2]}\over 
{(1+\sin\alpha\cos 2\psi)
-{1\over 2}(\sin\alpha +\cos 2\psi) \tr[g_2\sigma_2g_2^{-1}\sigma_2]}}
 \cr} \eqn\four$$
If we choose for $g_2$ the parametrization
 $g_2 = e^{i\gamma\sigma_2}e^{is\sigma_3}e^{i\beta\sigma_2}$,
and evaluate $\Gamma(g_2)$ in this coordinate system,
we finally get the lagrangian
$$\eqalign{&S =  -{k\over 2\pi}\int \d^2z \partial \psi{\bar\partial \psi}
\cr &~~~ +{k\over 2\pi}
\int \d^2z \left[\partial s {\bar\partial s} + \cos^2s\partial\rho
{\bar\partial\rho} + \sin^2s \partial\lambda{\bar\partial\lambda}
-{1\over 2}\cos (2s)(\partial\rho\overline\partial\lambda
-\overline\partial\rho\partial\lambda)\right]\cr
&~~~~~+{k\over \pi}\int \d^2z \left[
{(\cos 2\psi +\sin\alpha)
(\cos^2s\partial\rho-\sin^2s\partial\lambda)(\cos^2s{\bar\partial\rho}
+\sin^2s{\bar\partial\lambda}\over {(1 - \cos2\psi\cos2s)
 + \sin\alpha(\cos2\psi -  \cos2s)}}
\right] \cr
} \eqn\five$$
where $\rho = \gamma +\beta$ and $\lambda=\gamma-\beta$.
This can be rewritten
$$\eqalign{&S =  {k\over 2\pi}\int \d^2z \left[
-\partial \psi{\bar\partial \psi}
+ \partial s {\bar\partial s} + \partial\rho{\bar\partial\rho}
{{2\cos^2s\cos^2\psi(1 + \sin\alpha)}\over{(1- \cos2\psi\cos2s) +
 \sin\alpha(\cos2\psi -  \cos2s)}}\right.\cr &~~
+\partial\lambda{\bar\partial\lambda}
{{2\sin^2s\sin^2\psi(1-\sin\alpha)}\over
{(1- \cos2\psi\cos2s) +  \sin\alpha(\cos2\psi -  \cos2s)}}\cr
&~~\left.
+{1\over 2}(\partial\rho{\bar\partial\lambda} -
\partial\lambda{\bar\partial\rho
   })
{(\cos 2\psi-\cos 2s)+\sin\alpha(1-\cos 2s\cos 2\psi)\over
{(1- \cos2\psi\cos2s) +
 \sin\alpha(\cos2\psi -  \cos2s)}} \right]\cr} \eqn\bore$$
By
identifying this Lagrangian with the $\sigma$-model action of the
form
$$S = \int \d^2z (G_{MN}+B_{MN})\partial X_M{\bar\partial} X_N, \eqn\six$$
one can
read off the background space-time metric and antisymmetric tensor field.
The geometry in particular is described by a Lorentz signature
diagonal metric $G_{MN}$ where (with a
 common factor ${k\over 2\pi}$ removed)
$$\eqalign{&G_{\psi\psi}=-1\cr &  G_{ss}=1 \cr &
G_{\rho\rho}={{2\cos^2s\cos^2\psi(1 + \sin\alpha)}\over{(1- \cos2\psi\cos2s) +
 \sin\alpha(\cos2\psi -  \cos2s)}}\cr &
G_{\lambda\lambda}={{2\sin^2s\sin^2\psi(1-\sin\alpha)}\over
{(1- \cos2\psi\cos2s) +  \sin\alpha(\cos2\psi -  \cos2s)}}.\cr} \eqn\seven$$
The antisymmetric tensor field can similarly be read off from \bore.

\REF\buscher{T. H. Buscher, Phys. Lett. {\bf B201} (1988) 466.}
\REF\kir{E.B.Kiritsis, Mod. Phys.Lett. {\bf A6}, 2871 (1991).}
The dilaton background
can be derived either by following Buscher [\buscher, \kir]
and doing properly the gauge field
 integration
in \three\  or by imposing
the equations for the background fields derived by the string effective
 action, which are as we recall
$$R_{MN} + 2\nabla_M\nabla_N\phi - {1\over 4}H^2_{MN} =0.$$
Either way, the result is
$$\phi = -{1\over 2}\ln[(1- \cos2\psi\cos2s) +
 \sin\alpha(\cos2\psi -  \cos2s) ] .\eqn\eight$$

Now, we want to think of this world as a cosmological model,
with $\psi$ as the time parameter, running in the range
$0\leq \psi\leq \pi/2$.  At $\psi=0$, the universe begins from
a collapsed state, since $G_{\lambda\lambda}=0$.  At $\psi=\pi/2$,
the universe again collapses, since then $G_{\rho\rho}=0$.
Because of the factor of $k/2\pi$ that has been suppressed above,
the time scale for the expansion and recontraction is proportional
to $k$.  The maximum spatial extent of the universe is reached at
$\psi\sim \pi/4$ and is again of order $k$.

Now let us discuss the nature of the big bang singularity in the model.
First let us discuss why there is such a singularity.
The integrated form of the gauge transformation
law is
$$\eqalign{ g_1 & \to \exp({\epsilon \sigma_3})g_1\exp({\widetilde\epsilon}
\cos\alpha \sigma_3 +\epsilon\sin\alpha\sigma_3) \cr
            g_2 & \to \exp(i\widetilde\epsilon\sigma_2)g_2\exp(i\epsilon
\cos\alpha\sigma_2 -i{\widetilde\epsilon}\sin\alpha\sigma_2). \cr} \eqn\nucu$$
For simplicity let us take $\alpha=0$.
Setting $\epsilon=2\pi n$, $\widetilde \epsilon=2\pi m$, with $n$ and $m$
integers,  gives a $\IZ\times \IZ$ subgroup of the gauge group under
which $g_2$ is invariant.  At $\psi=0$, $g_1$ is invariant under the
subgroup with $n=-m$, and at $\psi=\pi/2$, $g_1$ is invariant under the
subgroup with $n=m$.  So the hypersurfaces $\psi=0$, $\psi=\pi/2$
are orbifold singularities of infinite order, each invariant under
a subgroup isomorphic to $\IZ$.
It is easy  to verify that $\psi=0$ and $\psi=\pi/2$ are
similarly orbifold
singularities for any $\alpha$ such that ${{\cos\alpha}/(1+\sin\alpha)}$
is a  rational number.   For other $\alpha$'s
$\psi=0,\pi/2$ are not quite orbifold singularities but points
that are arbitrarily close to being left invariant by suitable
elements of the gauge group
(found by approximating $\cos\alpha/(1+\sin\alpha)$ by  rational
numbers $n/m$).
This corresponds to a more subtle type of singularity in
the action of the gauge group
(possible only for actions of non-compact groups and therefore
possibly unfamiliar).

If $\psi=0$ and $s=0$ or $\psi=\pi/2$ and $s=\pi/2$, then the
pair $(g_1,g_2)$ is actually invariant under a continuous subgroup
of the gauge group
(for instance, $\widetilde\epsilon=\epsilon\cdot
{{\cos\alpha}/(1-\sin\alpha)}$
 for $\psi=s=0$).
Precisely at $\psi=s=0$ or $\psi=s=\pi/2$, the dilaton field $\phi$
blows up, according to equation \eight\ above; and other invariant
measures of the  physics similarly blow up at $\psi=s=0$ or
$\psi=s=\pi/2$.  This is in keeping with the fact that in constructions
of gauged WZW models, fixed points of continuous symmetry groups
typically give rise to space-time singularities; for instance, this
is the origin of the singularity of the two dimensional black hole
of [\witten].  Since the singularities at $\psi=s=0$ and $\psi=s=\pi/2$
are only present at one particular value of the ``time'' $\psi$,
they might appear as candidates for naked singularities violating
cosmic censorship.  What prevents this interpretation is precisely
that the whole universe collapses at the time when the naked singularity
would have appeared.  Whether naked singularities form from non-singular
initial data in other gauged WZW models is an interesting question
to which we hope to return.

\REF\hawking{S. Hawking and G.F.R.Ellis, {\it The Large Scale Structure of
Space-Time}, Cambridge University Press, 1973}
For $\psi=0$ or $\pi/2$ and generic $s$, one has not a curvature
singularity, but a kind of orbifold singularity as explained above.
This type of orbifold singularity appears in examples of Misner
and Taub-NUT; for a full explanation see the book
of Hawking and Ellis [\hawking, pp.170-178].
Just as in that example, it is possible to continue past $\psi=0$
or $\psi=\pi/2$; the continued metric is smooth (away from the
singularities noted above at special values of $s$) but has
closed timelike curves.   For instance, the continuation past
$\psi=0$ can be made by taking the
gauge condition to be not the one we used above but instead, for instance,
$$ g_1={1\over{1+x}}\left(\matrix{ 1 & 1\cr -x & 1\cr}\right).\eqn\hxss$$
The point of this is that the one invariant, $W=\Tr \sigma_3 g_1\sigma_3
g_1{}^{-1}$, is equal to \hfill\break
$2(1-x)/(1+ x)$.
For $x$ positive, one has $W\leq 2$,
just as with the gauge choice used above, but one can continue to
$W>2$ just by letting $x$ become negative.
The singularity that one might expect at $W=2$
is avoided with the gauge choice \hxss\ because that value of $W$
is attained
not for $g_1=1$ -- dangerous because of the orbifold phenomenon noted
above -- but for
$$ g_1=\left(\matrix{ 1 & 1 \cr 0 & 1\cr}\right),\eqn\huco$$
which is not an orbifold point.
\foot{The matrix in \huco\ cannot be distinguished from $g_1=1$ by any
continuous gauge invariant function, since it can be conjugated arbitrarily
close to $g_1=1$ by a gauge transformation.  This is another
phenomenon that is possible only for non-compact groups.}
Therefore,
with the gauge \hxss\ one can continue from $W<2$ to $W>2$ without
encountering a singularity at $W=2$ (except for $s=0$).

We can make all this explicit.
The gauge transformation \nucu\ from \idi\ to \hxss\ is achieved by
taking
$$\epsilon=-{{\cos\alpha}\over{1+\sin\alpha}}{\tilde\epsilon}
= {1\over 2}\ln{{\sin\psi}
\over {\cos\psi}} \qquad  x= {{\sin^2\psi}\over{\cos^2\psi}}$$
The corresponding change of coordinates is
$$\lambda\to \lambda'=\lambda-
 {{1+\sin\alpha}\over{\cos\alpha}}\ln{{\sin\psi}
\over{\cos\psi}}\qquad \psi \to x=  {{\sin^2\psi}\over{\cos^2\psi}}$$
In terms of these new coordinates the metric reads
$$\eqalign{\d {\mit l}^2
=& {1\over 4}{{\cos^6\psi}\over{\sin^2\psi}}\left[{{2\sin^2s
(1+\sin\alpha)}\over{\cos^2\psi((1- \cos2\psi\cos2s) +
 \sin\alpha(\cos2\psi -  \cos2s))}} -1\right] \d x^2\cr
&+\d s^2 +\d\lambda'{}^2 {{2\sin^2s\sin^2\psi(1-\sin\alpha)}\over
{(1- \cos2\psi\cos2s) +  \sin\alpha(\cos2\psi -  \cos2s)}}\cr
&+\d\lambda'\d x{{2\sin^2s\cos^2\psi\cos\alpha}\over{
(1- \cos2\psi\cos2s) + \sin\alpha(\cos2\psi -  \cos2s)}}\cr&
+\d\rho^2 {{2\cos^2s\cos^2\psi(1 + \sin\alpha)}\over{(1- \cos2\psi\cos2s) +
 \sin\alpha(\cos2\psi -  \cos2s)}}\cr}\eqn\fin$$
The main point is that
\fin\ is regular at $x=0$ and so continues without problem to negative
$x$.  At $x<0$, however, there are closed time-like curves.  Indeed,
the coefficient of $\d\lambda'{}^2$
vanishes at $x=0$ and is negative for $x<0$, so for $x<0$, the circle
corresponding to the periodicity $\lambda'\to\lambda'+2\pi$
is a closed timelike curve.  Hence -- as was anticipated above --
away from special values of $s$ the
big bang singularity in the model is not a curvature singularity,
but a singularity in the causal structure of the space-time.

{}From $x=\tan^2\psi$, we see that negative $x$ corresponds to imaginary
$\psi$.  Therefore, instead of the change of variables
by which we obtained \fin, we could have found the geometry
for $W>2$ (but not the smooth continuation from $W<2$ to $W>2$)
by setting $\psi=i\widetilde\psi$ in equation \seven.
\REF\giv{A.Giveon and M. Rocek, Generalized Duality in Curved
 String-Backgrounds, IASSNS-91/84}

{\bf Note Added}. A. Giveon has pointed out to us that our solution
can also be obtained by an $ O(2,2;R)$ transformation [\giv] of (a suitable
real form of) a product of two dimensional Lorentzian and Euclidean
black holes. A related observation was also made in [\hor].
\refout
 \end